\documentclass[aps,prl,twocolumn,superscriptaddress,floats,showpacs,a4paper,nobibnotes]{revtex4}
\usepackage{newfloat}
\usepackage{latexsym}
\usepackage{dcolumn}
\usepackage{graphicx}
\usepackage{amssymb}
\usepackage{amsmath}
\usepackage{float}
\usepackage{hyperref}
\hypersetup{colorlinks,linkcolor=blue,citecolor=blue,urlcolor=blue}
\usepackage[left=18mm,right=18mm,top=25mm,bottom=25mm]{geometry}
\usepackage{mhchem}
\usepackage{bm}

\begin{document}

\title{Disorder induced magnetoresistance in a two dimensional electron system}

\author{Jinglei Ping}
\altaffiliation{Current address:  Department of Physics and Astronomy, University of Pennsylvania, 209 South 33rd Street, Philadelphia, PA 19104-6396, USA}
\affiliation{Center for Nanophysics and Advanced Materials, University of Maryland, College Park, Maryland 20742-4111, USA}
\affiliation{School of Physics, Monash University, Victoria 3800, Australia}

\author{Indra Yudhistira}
\affiliation{Graphene Research Centre and Department of Physics, National University of Singapore, 2 Science Drive 3, 117551, Singapore}

\author{Navneeth Ramakrishnan}
\affiliation{Graphene Research Centre and Department of Physics, National University of Singapore, 2 Science Drive 3, 117551, Singapore}

\author{Sungjae Cho}
\altaffiliation{Current address: Department of Physics, Frederick Seitz Materials Research Laboratory, University of Illinois, Urbana, Illinois 61801, USA}
\affiliation{Center for Nanophysics and Advanced Materials, University of Maryland, College Park, Maryland 20742-4111, USA}

\author{Shaffique Adam}
\email{shaffique.adam@yale-nus.edu.sg}
\affiliation{Graphene Research Centre and Department of Physics, National University of Singapore, 2 Science Drive 3, 117551, Singapore}
\affiliation{Yale-NUS College, 6 College Avenue East, 138614, Singapore}

\author{Michael S. Fuhrer}
\email{michael.fuhrer@monash.edu}
\affiliation{Center for Nanophysics and Advanced Materials, University of Maryland, College Park, Maryland 20742-4111, USA}
\affiliation{School of Physics, Monash University, Victoria 3800, Australia}
        
\date{\today}

\begin{abstract}
We predict and demonstrate that a disorder induced carrier density inhomogeneity causes magnetoresistance (MR) in a two-dimensional electron system. Our experiments on graphene show a quadratic MR persisting far from the charge neutrality point. Effective medium calculations show that for charged impurity disorder, the low-field MR is a universal function of the ratio of carrier density to fluctuations in carrier density, a power-law when this ratio is large, in excellent agreement with experiment. The MR is generic and should occur in other materials with large carrier density inhomogeneity.
\end{abstract}
\pacs{}
\maketitle

{\it Introduction -- }  The classical magnetoresistance of a material arises when the Lorentz force caused by an applied magnetic field has a component acting against the direction of electron motion thereby decreasing the conductivity of an electronic material.  This property has long been of interest both as a tool to probe the fundamental properties of an electronic material (such as the topology of the electron bands)~\cite{kn:ashcroft1976} and also for technological applications such as its use in magnetic memory read-heads~\cite{kn:nickel1995}.  A well known result is that single electronic bands (in systems with a spatially homogenous carrier density) will have no magnetoresistance, while the presence of two or more electronic bands with different carrier mobilities readily gives rise to a classical magnetoresistance. This classical effect is different from weak localization~\cite{kn:altshuler1980} (a quantum interference effect present at low temperatures) or Abrikosov's quantum magnetoresistance~\cite{kn:abrikosov1998} (that occurs in gapless semiconductors in the high field Landau quantized regime).

Graphene is an example of a material with more than one electronic band.  This single-atom-thick sheet of carbon comprises an electron band and a hole band each with a linear dispersion that touch at a topologically protected Dirac point~\cite{kn:neto2009,kn:dassarma2011a}.  It was shown by Ref.~\cite{kn:hwang2006e}  that if both bands are occupied, one then expects a classical magnetoresistance even if the electron and hole bands have the same electronic mobility (this is by virtue of the Lorentz force being of opposite sign for the electron and hole carriers).  While this two-channel model has been reasonably successful at modeling the density dependence of graphene magnetotransport at fixed magnetic field, it is unable to quantitatively explain the magnetic field dependence at fixed carrier density.  In this context, Ref.~\cite{kn:guttal2005} and Ref.~\cite{kn:tiwari2009} developed an effective medium approximation where the planar landscape is broken up into electron regions and hole regions with different area fractions.  Assuming that all electron regions had a uniform conductivity $\sigma_e$ and all hole regions had a uniform conductivity $\sigma_h$, they could calculate the magnetoresistance of this effective medium.  While this description works well at the charge neutrality point (where the electrons and holes regions have equal area fractions), it fails to adequately describe the experiments away from this symmetry point.  We refer the reader to Ref.~\cite{kn:cho2008} for a detailed discussion of these earlier theoretical and experimental results.  However, we note that both the two-channel model of Hwang {\it et al.} and the effective medium calculation of Stroud and collaborators both predict that the magnetoresistance should vanish away from the Dirac point when only a single band is occupied.  By contrast, in this work we discuss a carrier density inhomogeneity contribution to the magnetoresistance that persists away from the Dirac point and exists even if only one electronic band is occupied.  For concreteness we focus our discussion on graphene which has a linear dispersion, but the mechanism itself does not rely on the linear dispersion and should therefore be observable in other materials with large spatial inhomogeneity in the carrier density distribution.

The idea of a disorder induced magnetoresistance is not new. While working on silver chalcogenides, Parish and Littlewood~\cite{kn:parish2003,kn:parish2005} predicted such an effect by mapping the problem onto a random resistor network and solving it numerically. They focussed on the high magnetic field regime and found a linear magnetoresistance, but were otherwise unable to make quantitative comparisons with experiments. By contrast, in this work, we use an effective medium theory to study the low field regime and make quantitative predictions that are then compared to experimental results.

\begin{figure}
	\includegraphics[angle = 270, width=3.2in]{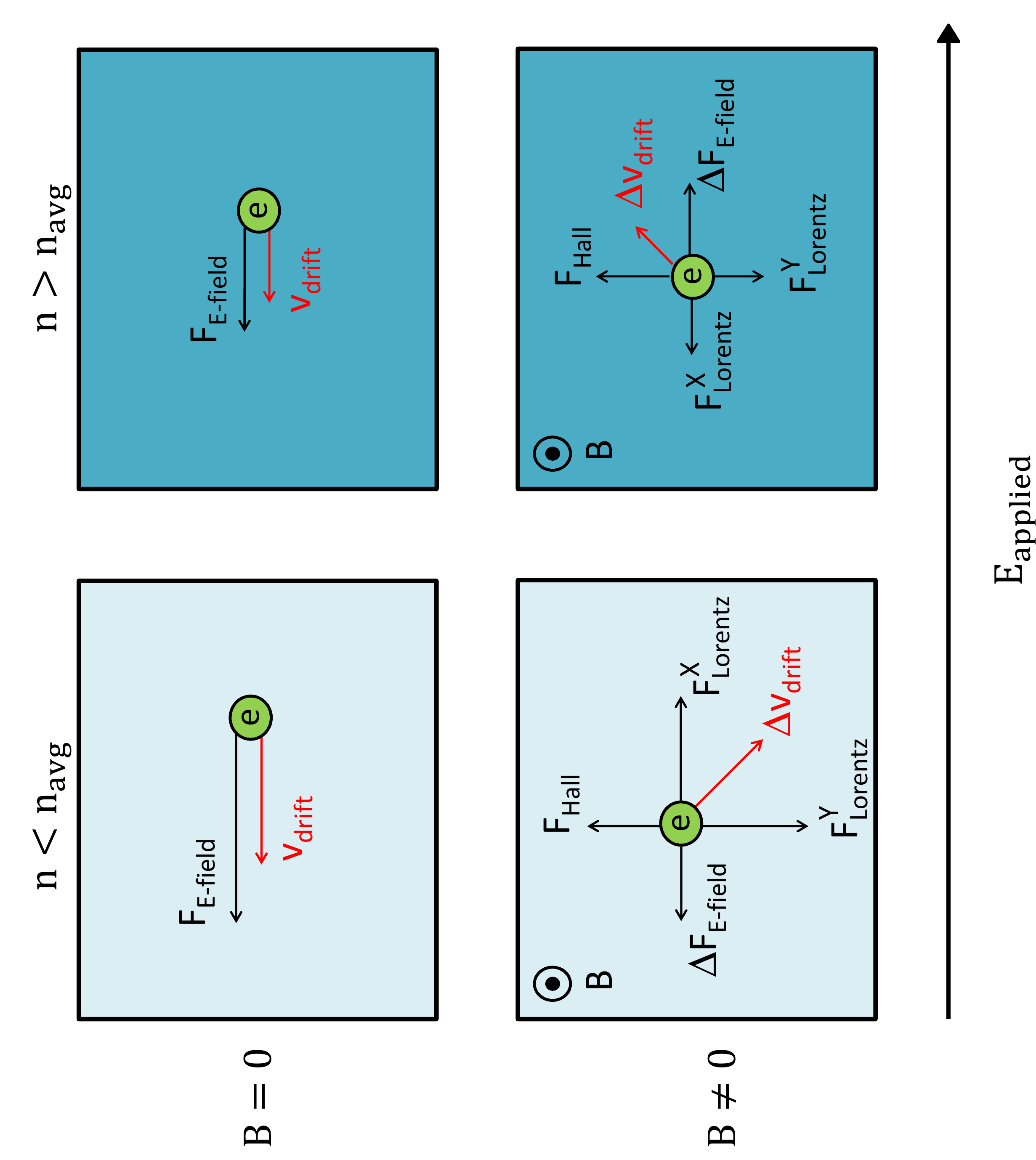}
	\caption{\label{Fig:fig1}(Color online) The microscopic origin of magnetoresistance in system with inhomogeneous carrier density: The upper panels show the forces (excluding the drag force) and drift velocities in a sample with two regions of different charge carrier concentrations when there is no magnetic field. The bottom panels illustrate all the new forces that appear in the presence of a magnetic field. The change in drift velocity $\Delta v_{\rm drift}$ is against the direction of motion of the electron in both regions, giving rise to magnetoresistance.
}
\end{figure}

The basic mechanism of inhomogeneity induced magnetoresistance can be seen in Fig.~\ref{Fig:fig1}.  We schematically show two regions where the local carrier densities (and local conductivities) are smaller and larger than the bulk average respectively.  In both regions the Hall field is the same and perpendicular to both the applied electric field and the applied magnetic field.  The magnetic field results in a Lorentz force acting on the charge carriers as well as an adjustment of the local electric fields such that current remains conserved in the direction of the applied electric field. Requiring that the net adjustment of the electric field over the sample be zero results in a reduction of the drift velocity in both the high and low conductivity regions in the presence of magnetic field. Below, we develop a full effective medium theory that quantitatively captures this effect.

\begin{figure*}
\includegraphics[width=6.4in]{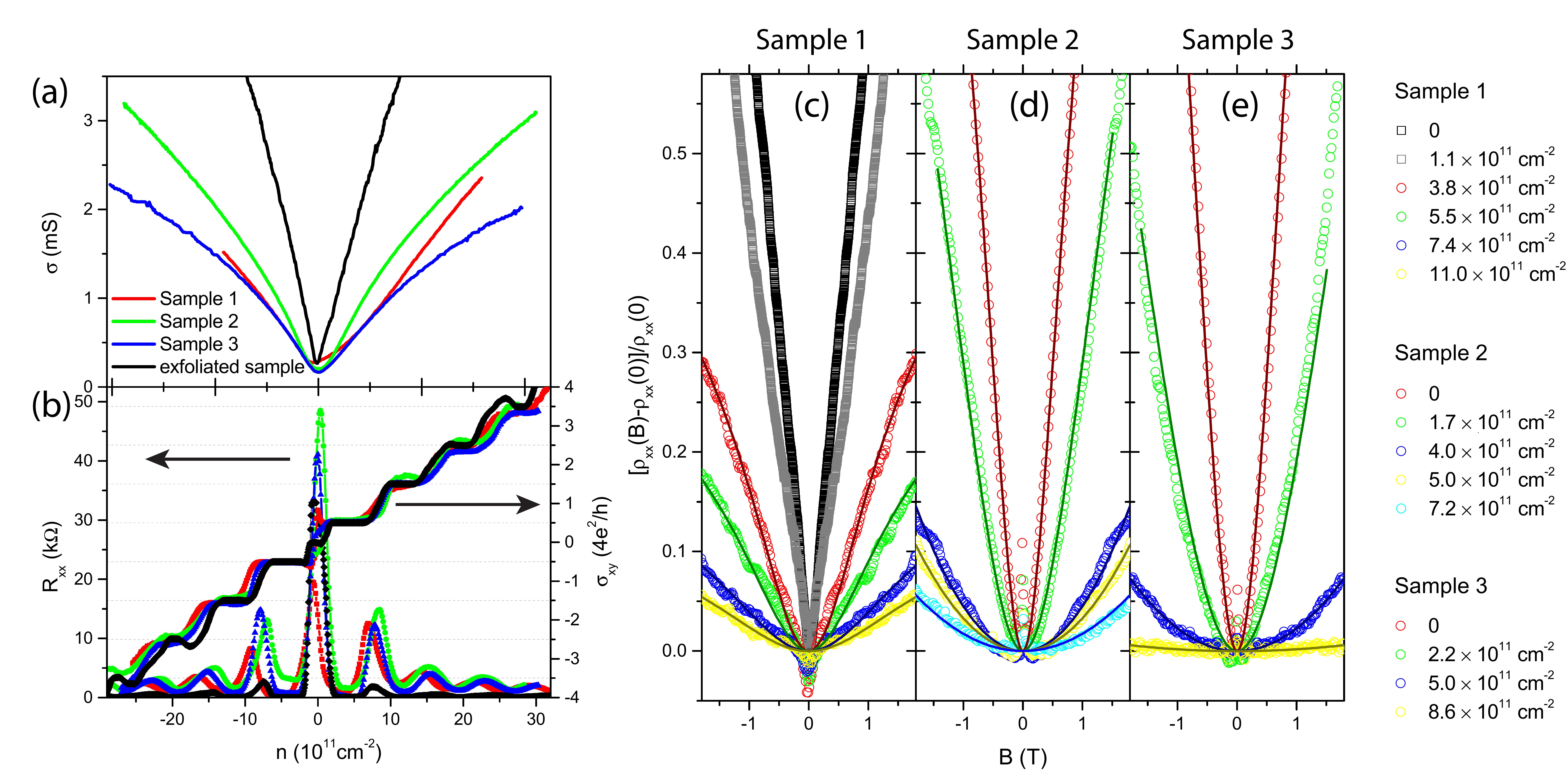}
\caption{\label{Fig:fig2} (Color online) Summary of the experimental results. (a) Conductivity as a function of back-gate voltage at zero magnetic field for CVD-grown samples 1, 2 and 3, and the exfoliated graphene sample from ~\cite{kn:cho2008}. (b)Longitudinal resistance $R_{xx}$ and Hall conductivity $\sigma_{xy}$ as a function of back gate voltage for CVD-grown samples 1,2 and 3 and exfoliated sample at $T=4.2~{\rm K}$ and $B = 8~{\rm T}$. (c) Low field magnetoresistance of samples 1 (left), 2 (middle) and 3 (right). Data are open symbols taken at gate voltage indicated in legend. Solid lines are fits described in the text.}
\end{figure*}

{\it Experimental procedure -- }   We measure the magnetotransport in single-crystal graphene synthesized by chemical vapor deposition on Pt foil~\cite{kn:gao2012,kn:ping2013}. The three CVD-grown samples 1, 2, and 3 were prepared at temperatures of 1000~$^\circ${\rm C}, 950~$^\circ${\rm C} and 900~$^\circ${\rm C}, and hydrogen mass flow rates of 700 sccm, 500 sccm and 380 sccm, respectively. The synthesized graphene is then coated with PMMA with a spinning speed of 2000 rpm and then transferred to a 300 nm SiO$_2$ on Si substrate by electrolysis method~\cite{kn:gao2012}.  Electron-beam lithography using poly(methyl methacrylate) resist is used to establish Cr/Au contacts via liftoff and again to define the graphene in a Hall-bar geometry via oxygen plasma etching.  All three devices have the same geometry shown in the inset of Fig.~\ref{Fig:fig2}b. 

As discussed in detail elsewhere~\cite{kn:ping2013}, the differences between the Raman spectra of the three samples suggest the presence of nanocrystalline carbon impurities on the continuous crystalline graphene layer; sample 1 has the greatest concentration of impurities and sample 3 has the least. These impurities do not correlate with mobility, and so for the purposes of this work, the samples simply have varying amounts of disorder. 

Fig.~\ref{Fig:fig2}a shows the zero-field conductivity as a function of back gate induced carrier density $n_0$ for samples 1, 2 and 3. We observe the typical approximately linear dependence of conductivity as a function of carrier density~\cite{kn:dassarma2011a}, and from the data we can extract the charge-impurity limited mobility $\mu$ of the three samples as 8,300~{\rm cm}$^2$/{\rm Vs}, 8,100~{\rm cm}$^2$/{\rm Vs}, and 10,700~{\rm cm}$^2$/{\rm Vs} for samples 1, 2 and 3 respectively (see supplementary material for more details).  These values of mobility are among the highest for CVD-grown graphene transferred to \ce{SiO2} method. Also shown in Fig.~\ref{Fig:fig2}a  are data from an exfoliated graphene sample~\cite{kn:cho2008}, which has charge-impurity limited mobility $\mu$ of 18,200~{\rm cm}$^{2}$/{\rm Vs}.       

We also experimentally measure the minimum conductivity $\sigma_{\rm min}$, and use this to extract the disorder induced carrier density fluctuations $n_{\rm rms}$ using $\sigma_{\rm min} = n_{\rm rms} e \mu/\sqrt{3}$.  The values for $n_{\rm rms}$  extracted this way are between $20\%$ and $40\%$ lower than what one would expect from the self-consistent theory for graphene transport~\cite{kn:adam2007a} that is normally~\cite{kn:dassarma2011a}  used to understand the graphene minimum conductivity.  We attribute this discrepancy to the non-perfect transmission across p-n junctions separating the electron and hole regions or due to additional scattering by the nanocystaline grain boundaries.  As will become clearer later, we parameterize our data as a function of the ratio $n_0/n_{\rm rms}$, where both the average carrier density $n_0$ and the density fluctuations $n_{\rm rms}$ are measured independently.  

\begin{figure}
\includegraphics[width=2.8in]{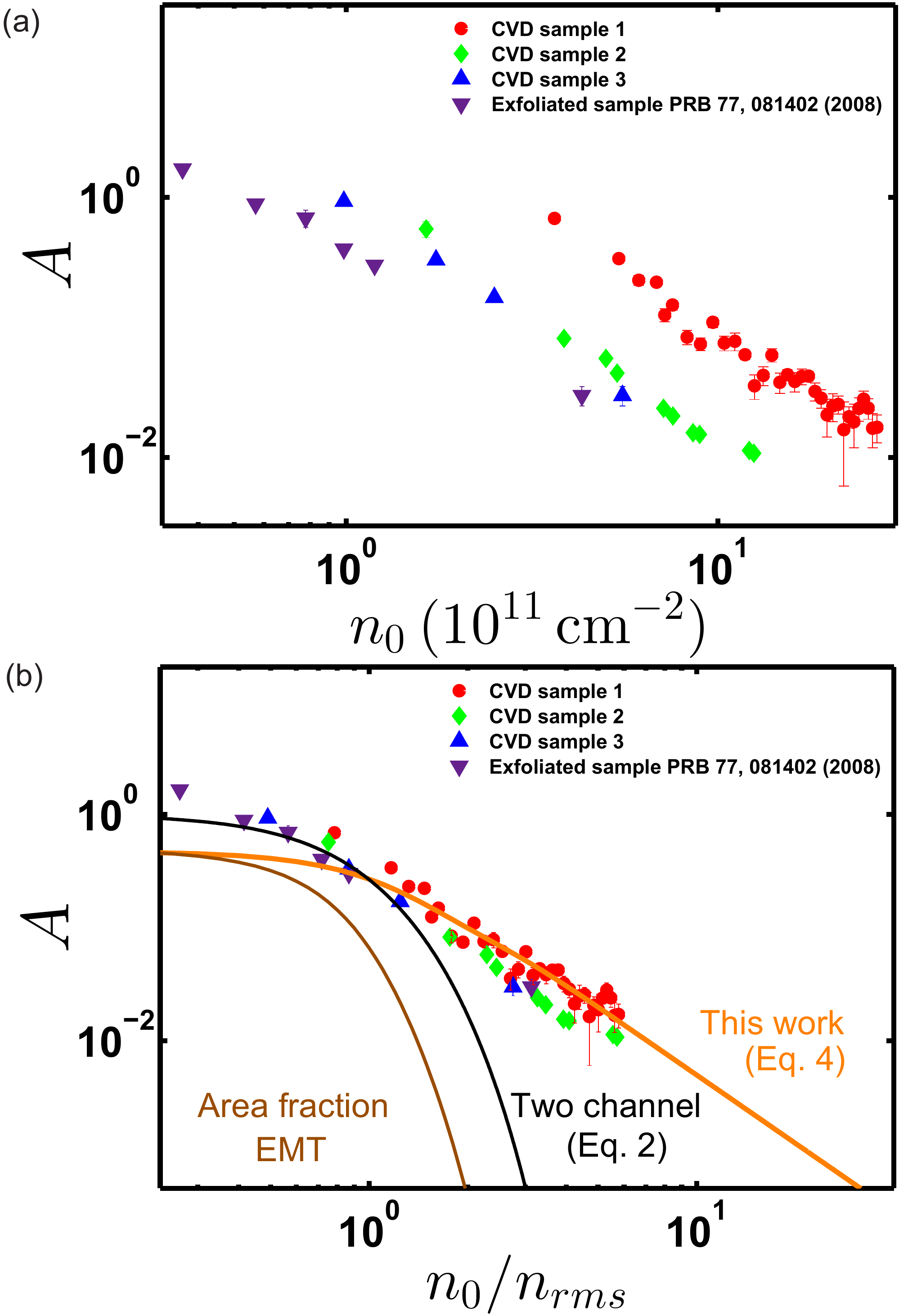}
\caption{\label{Fig:fig3}(Color online) Theoretical and experimental results for the dependence of the coefficient of quadratic magnetoresistance (a) plotted as a function of the carrier density $n_0$ and (b) as a function of the ratio between carrier density $n_0$ and carrier density fluctuations $n_{\rm rms}$.  The magnetoresistance in the earlier theoretical models including the two channel model by \protect{Ref.~\cite{kn:hwang2006e}} and the area fraction effective medium theory by \protect{Ref.~\cite{kn:tiwari2009}}. In both earlier models, the magnetoresistance vanishes quickly once $n_0 > n_{\rm rms}$.  By contrast, the magnetoresistance discussed in this work persists away from the Dirac point.  For $n >> n_{\rm. rms}$, we find both theoretically and experimentally a power law dependance: $A = (1/2)(n_0/n_{\rm rms})^{-2}$.} 
\end{figure}

Fig.~\ref{Fig:fig2}b shows the longitudinal resistance $R_{xx}$ and Hall conductivity $\sigma_{xy}$ as a function of back gate voltage for all three samples at $T=4.2~{\rm K}$ and $B = 8~{\rm T}$.  The data clearly shows Subnikov-de Haas oscillations and quantum Hall plateaus with $\sigma_{xy}  = 4(n +1/2)e^2/h$ (where the factor 1/2 is the  fingerprint of the $\pi$ Berry's phase in monolayer graphene)~\cite{kn:novoselov2005,kn:zhang2005,kn:wu2007}. Fig.~\ref{Fig:fig2} shows the magnetoresistance. At high magnetic field and low carrier densities, we sometimes observe a linear magnetoresistance. However, for sufficiently low magnetic fields, we always observe a quadratic magnetoresistance, where for different values of carrier density $n_0$ and density fluctuation $n_{\rm rms}$, we can fit our data to 
\begin{equation}
\label{Aeqn}
\rho_{xx}(B) = \rho_{xx}(B=0) \left[ 1 + A (\mu B)^2 \right]
\end{equation}
and extract the dimensionless coefficient $A[n_0, n_{\rm rms}]$ from our data. We can also fit the data over a larger range of B using the phenomenological formula of Ref.~\cite{kn:cho2008}, $\rho_{xx}(B) = \rho_{xx}(0)\left[1-\alpha+\frac{\alpha}{\sqrt{1+\frac{2A(\mu B)^{2}}{\alpha}}}\right]^{-1}$, where $\alpha$ is a fitting parameter. Notice that for $\mu B<<1$, this phenomenological expression gives the same value for the quadratic magnetoresistance as Eq.~\ref{Aeqn}. These fits are shown in Fig.~\ref{Fig:fig2}.  We find experimentally that $A[n_0, n_{\rm rms}]$ scales as a function of the ratio $n_0/n_{\rm rms}$ and that for large $n_0/n_{\rm rms}$ it follows a power law $A \sim (n_0/n_{\rm rms})^{-2}$.  Our experimental results shown in Fig.~\ref{Fig:fig3} suggest that the magnetoresistance persists far away from the Dirac point (i.e. $n_0 > n_{\rm rms}$)  and is caused by carrier density inhomogeneity and not due to the presence of both electrons and holes close to the Dirac point.        

 {\it Theoretical analysis -- }   The starting point for this analysis is to assume that the carrier density $n$ is Gaussian distributed centered at an average carrier density $n_0$ with an {\rm rms} fluctuation given by $n_{\rm rms}$ (we denote this distribution as $P[n, n_0, n_{\rm rms}]$.  For the specific case of graphene, Ref.~\cite{kn:adam2009} justified theoretically the use of a Gaussian distribution, and at least close to charge neutrality, this has been seen in several experiments starting with Ref.~\cite{kn:martin2008}.  In this context ${\rm sign}(n_0) = \pm 1$ represents the electron and hole bands respectively.  For the case where charged impurities dominate the transport properties, knowing the impurity concentration $n_{\rm imp}$ and the distance $d$ away from the graphene sheet, one can calculate  $n_{\rm rms}$ both analytically~\cite{kn:adam2007a} (using the self-consistent approximation) and numerically~\cite{kn:rossi2008} (using the mesoscopic density functional approach). The carrier mobility $\mu$ is calculated using the semi-classical Boltzmann transport theory~\cite{kn:dassarma2011a}.  In what follows, for simplicity, we assume that $\mu$ is density independent, and we show in the supplementary material that the weak density dependence of the carrier mobility hardly changes any of our results.  Since $n_0$ is controllably tuned by a back gate, and the parameters $n_{\rm imp}$ and $d$ can be obtained from the conductivity at zero magnetic field, all the parameters used in our theory for magnetoresistance can be fixed by measurements done before applying a magnetic field.  

Before discussing our inhomogeniety induced magnetoresistance, we first briefly comment on previous theories for graphene magnetoresistance in the context of our framework.  For a single channel model, the transport in the presence of a magnetic field is given by~\cite{kn:ashcroft1976} 
\begin{equation}
\label{Eq:onechannel}
\sigma(B) = \frac{|n_0| e \mu}{1 + (\mu B)^2} \left( \begin{array}{cc}  1  & +{\rm sign}(n_0) \mu B \\ -{\rm sign}(n_0) \mu B & 1 \end{array} \right),  
\end{equation}
\noindent where it is easy to verify that there is no magnetoresistance i.e. $\rho_{xx}(B) = \sigma_{xx} /(\sigma_{xx}^2 + \sigma_{xy}^2) = \rho_{xx}(0)$.  The two channel model~\cite{kn:hwang2006e} assumes that the total conductivity is the sum of the electron and hole channels, $\sigma_{xx} = \sigma_{xx}^e +  \sigma_{xx}^h$, and similarly for the transverse conductivity, $\sigma_{xy} =  \sigma_{xy}^e +  \sigma_{xy}^h$.  Defining $ = n_0/n_{\rm rms}$, a straightforward calculation gives the quadratic coefficient of the magnetoresistance (see definition above) as
\begin{equation}
A \left[ \eta = \frac{n_0}{n_{\rm rms}} \right] = 1 - \left( \frac{\eta \sqrt{2 \pi}}{2e^{-\eta^2/2} + \eta \sqrt{2 \pi} {\rm Erf}(\eta/\sqrt{2})}  \right)^2
\label{Eq:twochannel}
\end{equation}

Here ${\rm Erf}(x)$ is the error function~\cite{kn:gradshteyn1994}.  Notice that $A[\eta]$ is independent of the carrier mobility $\mu$ and depends only on the ratio of the carrier density and density fluctuation.  This remains true so long as $\mu$ is independent of carrier density, and in this sense $A[\eta]$ becomes a universal function (where the different theoretical models each give a different functional form for $A[\eta]$). The two-channel result Eq.~\ref{Eq:twochannel} is shown in Fig.~\ref{Fig:fig3}; it has the value $A=1$ at the Dirac point, stays roughly constant for $n_0 < n_{\rm rms}$, and then rapidly decreases for $n>n_{\rm rms}$ as the second channel becomes depopulated. The inadequacy of this model to explain experimental data led Ref.~\cite{kn:tiwari2009} to develop an area-fraction effective medium theory.  This model assumes that there are electron regions with area-fraction $f_e$ and conductivity $\sigma_e = n_e e \mu$, and hole regions with $f_h$ and $\sigma_h = n_h e \mu$. The effective medium conductivity tensor $\sigma_{\rm EMT}$ is obtained by solving $\sum_{i = e,h} f_i \delta \sigma_i (\openone_2 - \Gamma \delta \sigma_i)^{-1} = 0$, where the shorthand notation $\delta \sigma_i = \sigma_i - \sigma_{\rm EMT}$ is used. In the case where the electron and hole puddles can be assumed to be nearly circular, the depolarization tensor $\Gamma =  -\openone_2 /(2 \sigma_{\rm EMT}^{xx})$ takes a simple scalar form (see Ref.~\cite{kn:stroud1975} for details).  In this case, a remarkable result~\cite{kn:guttal2005} is that when $n_0 = 0$ (and hence $f_e = f_h$), the magnetoresistance is given by  $\rho_{xx}(B)  = \rho_{xx} (0) \sqrt{1 + (\mu B)^2}$.  Since the self-consistent theory~\cite{kn:adam2007a} gives $\rho_{xx}(0) = \sqrt{3}/(n_{\rm rms} e \mu)$ and $\mu [{\rm m}^2/{\rm Vs}] \approx 50/(n_{\rm imp}[10^{10} {\rm cm}^{-2}])$, the full magnetoresistance at the Dirac point is completely specified.  In particular, we have $A[0] = 1/2$.  This model can be solved numerically away from the Dirac point, and the results are shown in Fig.~\ref{Fig:fig3}.  Notice again that for $n_0>n_{\rm rms}$, the area fraction of the hole channel vanishes and the magnetoresistance drops rapidly. 

The inadequacy of the two-channel model is that it does not account for the spatial inhomogeneity of the carrier density, and the inadequacy of the area-fraction EMT is that although it allows for two dimensional space to broken up into regions of electron and hole puddles, all electron and hole regions are assumed to be uniform.  What is required is an effective medium approach with a continuous distribution of carrier density (similar to what has been developed in Refs.~\cite{kn:rossi2008b,kn:fogler2008b} for transport in zero-magnetic field).  Using the form of the depolarization tensor derived in Ref.~\cite{kn:stroud1975}, for our system, we can derive a set of coupled equations
\begin{widetext}
\begin{subequations}
\label{Eq:main}
\begin{equation}
\int dn P[n,n_0, n_{\rm rms}]   \frac{\sigma_{xx}^2[n] - (\sigma^{\rm EMT}_{xx})^2 + (\sigma^{\rm EMT}_{xy} - \sigma_{xy}[n] )^2}{(\sigma^{\rm EMT}_{xx} + \sigma_{xx}[n] )^2 +  (\sigma^{\rm EMT}_{xy} - \sigma_{xy}[n] )^2} = 0
\end{equation}
\begin{equation}
\int dn P[n,n_0, n_{\rm rms}]  \frac{\sigma_{xy}[n] - \sigma^{\rm EMT}_{xy}}{(\sigma^{\rm EMT}_{xx} + \sigma_{xx}[n] )^2 +  (\sigma^{\rm EMT}_{xy} - \sigma_{xy}[n] )^2}=0.
\end{equation}
 \end{subequations}  
\end{widetext}
It is understood from Eq.~\ref{Eq:main} that $\sigma_{xx}[n]$ and $\sigma_{xy}[n]$ are obtained from some homogenous density model (such as Eq.~\ref{Eq:onechannel}) and then these coupled integral equations give the correct averaging over the density inhomogeneity.  One can verify that for $B=0$, we get $\sigma^{\rm EMT}_{xy} =0$, and the equation for $\sigma^{\rm EMT}_{xx}$ reproduces the zero-magnetic field effective medium theory results of ~\cite{kn:adam2009} and ~\cite{kn:rossi2008b}.  Moreover $\sigma^{\rm EMT}_{xy} =0$ also for $n_0 = 0$, and a numerical solution of the $\sigma^{\rm EMT}_{xx}$ gives results very close to $\rho_{xx}(B)  = \rho_{xx} (0) \sqrt{1 + (\mu B)^2}$ (although, technically, it need not have given the same result since our model allows for carrier density inhomogeneity inside each puddle).  We emphasize that Eq.~\ref{Eq:main} can be solved with any model for the density profile $P[n,n_0,n_{rms}]$ and scaterring potential as input for $\mu$.  In the simplified case where $\mu$ is independent of density, e.g. for charged impurity scattering, Eq.~\ref{Eq:main} simplifies considerably and the normalized magnetoresitance $\rho_{xx}(B)/\rho_{xx}(0)$ depends only on the ratios $n_{0}/n_{\rm rms}$ and $\mu B$. In this case, both $\sigma_{xx}^{\rm EMT}$ and $\sigma_{xy}^{\rm EMT}$ can be written in term of dimensionless coefficient $y_1$ and $y_2$ as
\begin{subequations}
\begin{equation}
\sigma_{xx}^{\rm EMT} = y_1~n_{\rm rms} e \mu/(1+ (\mu B)^2)
\end{equation}
\begin{equation}
\sigma_{xy}^{\rm EMT} = y_2~n_{\rm rms} e \mu^2 B/(1+ (\mu B)^2),
\end{equation}
\end{subequations}
where $y_{1}=y_{1}\left[\frac{n_{0}}{n_{{\rm rms}}},\mu B\right]$ and $y_{2}=y_{2}\left[\frac{n_{0}}{n_{{\rm rms}}},\mu B\right]$ are computed in the supplementary material. The coefficient of quadratic magnetoresistance obtained by solving these equations numerically has been plotted in Fig.~\ref{Fig:fig3}.

It is important to notice that our inhomogenous carrier density driven magnetoresistance persists far away from the Dirac point, and is not specific to the linear dispersion of graphene. Generally, there is remarkable agreement between the theoretical and experimental results presented here. As we explain in the supplementary material, the discrepancy close to the Dirac point (for small values of $\eta$) can be directly traced to the overestimating of $\sigma_{\rm min}$ in the self-consistent theory and therefore a difference between the theoretical and experimental values used for $n_{\rm rms}$ close to the Dirac point.

In summary we have shown both theoretical and experimental results for an inhomogeneity induced quadratic magnetoresistance that scales as a power law of the ratio $n_0/n_{\rm rms}$.  While we focused on the case of charged impurities in graphene,  the mechanism itself requires only spatial fluctuations in the carrier density and should therefore be observable in other systems.

{\it Acknowledgement:}   The experimental work was supported by the University of Maryland NSF-MRSEC under Grant No. DMR 05-20471 and the US ONR MURI program. MSF acknowledges support from an ARC Laureate fellowship. The theoretical work in Singapore was supported by the National Research Foundation Singapore under its Fellowship program (NRF-NRFF2012-01).

\bibliography{shaffiquebib}

\clearpage
\onecolumngrid
\vspace*{0.4cm}
\begin{center}
{\large\bf SUPPLEMENTARY MATERIAL}\\
\vspace*{0.4cm}
\end{center}

\renewcommand{\thetable}{S\arabic{table}}
\renewcommand{\thefigure}{S\arabic{figure}}
\renewcommand{\thesection}{S\arabic{section}}
\renewcommand{\thesubsection}{S\arabic{subsection}}
\renewcommand{\theequation}{S\arabic{equation}}

\setcounter{secnumdepth}{3}

\twocolumngrid
\setcounter{equation}{0}
\setcounter{figure}{0}

\subsection{Conductivity fitting at zero magnetic field}

\begin{figure}[h]
\includegraphics[width=3.2in]{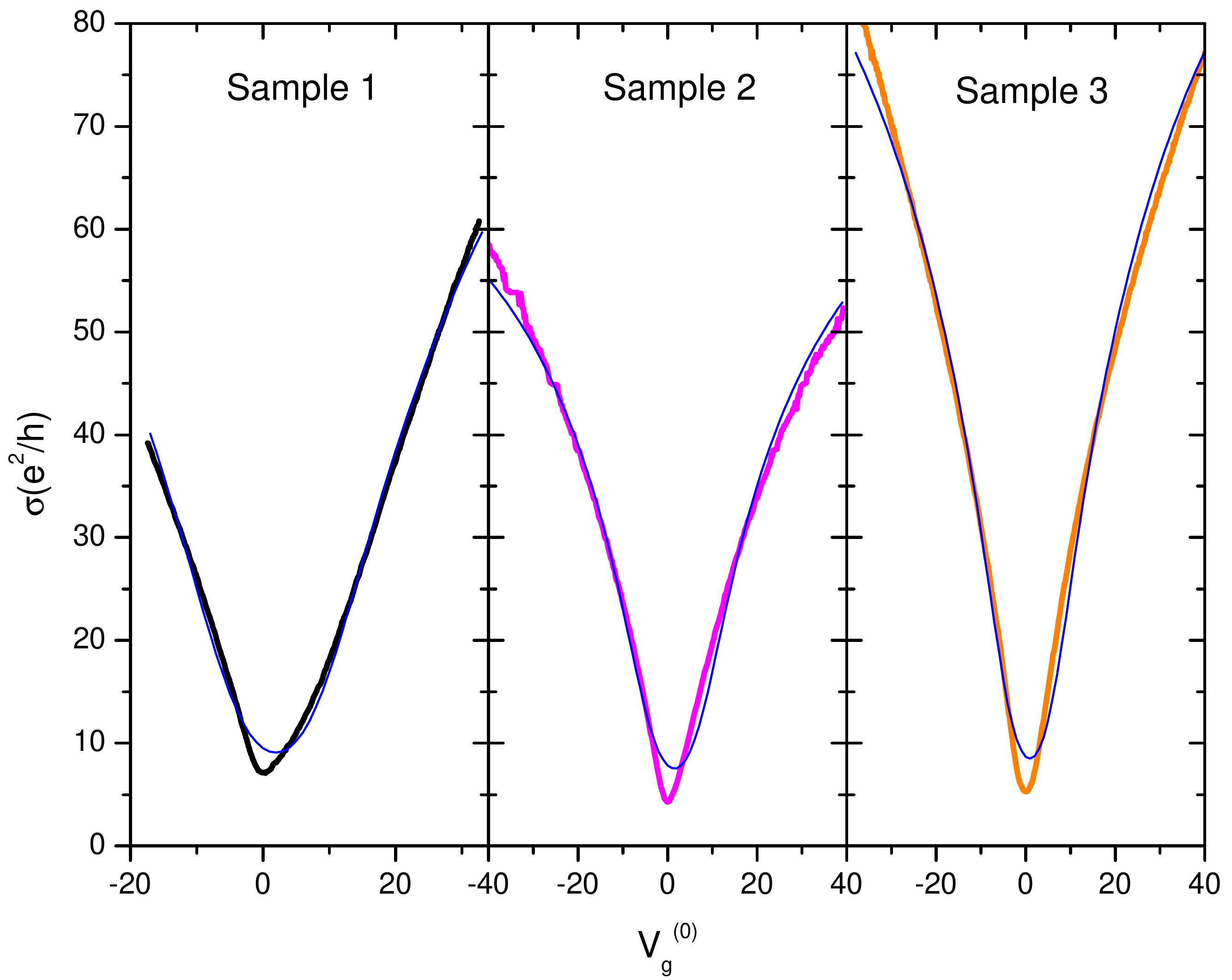}
\caption{\label{Fig:S1}
(Color online) Fitting of effective medium theory conductivity to experimental results at zero magnetic field.}
\end{figure}

Following a standard procedure~\cite{ref1}, we fit conductivity of experiment to conductivity of effective medium theory at zero magnetic field as shown in Fig.~\ref{Fig:S1}. There are three parameters used in the fitting, which are short-range scatterers $\sigma_{\mathrm{s}}$, charged impurity density $n_{\mathrm{imp}}$ and the distance between graphene and the substrate $d$. These parameters enter the EMT equations through the RPA-Boltzmann conductivity $\sigma_{\mathrm{B}}[\sigma_\mathrm{{s}},n_{\mathrm{imp}},d,r_{\mathrm{s}}]$, where $r_{\mathrm{s}}=0.8$ is used for graphene on \ce{SiO2}. The RPA-Boltzmann conductivity is used in the zero magnetic field EMT equations
\begin{equation}
\int_{-\infty}^{\infty}dn\, P[n,n_{0},n_{\rm rms}]\frac{\sigma_{\mathrm{B}}-\sigma_{\mathrm{EMT}}}{\sigma_{\mathrm{B}}+\sigma_{\mathrm{EMT}}}=0
\end{equation}
where $P[n,n_{0},n{\mathrm{rms}}]$ is a Gaussian distribution centered at an average carrier density $n_0$ with an {\rm rms} fluctuation given by $n_{\rm rms}$. The fitting parameters for the three samples are shown in Table~\ref{tab:fitparam}.

\begin{table}[H]
\caption{\label{tab:fitparam}Fitting parameters of conductivity $\sigma$ vs gate voltage $V_{\mathrm{g}}$
of experiment with conductivity of effective medium theory at zero
magnetic field. The parameters short-range scatterers $\sigma_{s}$,
charged impurity density $n_{\mathrm{imp}}$ and the distance between
graphene and the substrate $d$ are used in the fitting.}

\centering{}%
\begin{ruledtabular}
\begin{tabular}{cccc} 
 & $\sigma_{\mathrm{s}}$ (e$^{2}$/h) & $n_{\mathrm{imp}}$ ($10^{10}$ cm$^{-2}$) & $d$ (nm)\tabularnewline
\hline\tabularnewline
CVD Sample 1 & 137.4 & 58.4 & 0.52\tabularnewline 
CVD Sample 2 & 79.1 & 59.6 & 0.80\tabularnewline 
CVD Sample 3 & 121.9 & 45.3 & 0.72\tabularnewline
Exfoliated Sample & 729.0 & 26.5 & 0.14\tabularnewline 

\end{tabular}
\end{ruledtabular}
\end{table}

The charge-impurity limited mobility of the three CVD-grown graphene samples are then calculated to be 8,300~{\rm cm}$^2$/{\rm Vs}, 8,100~{\rm cm}$^2$/{\rm Vs}, and 10,700~{\rm cm}$^2$/{\rm Vs} for samples 1, 2, and 3 respectively and 18,200~{\rm cm}$^{2}$/{\rm Vs} for exfoliated graphene. The charged impurity limited conductivity seems uncorrelated with the island like impurities that are observed in an optical image. See Ref. ~\cite{ref2, ref3} for details about the island like impurities.

\subsection{Effective Medium Theory for constant mobility}
\begin{figure}[H]
\includegraphics[width=3.43in]{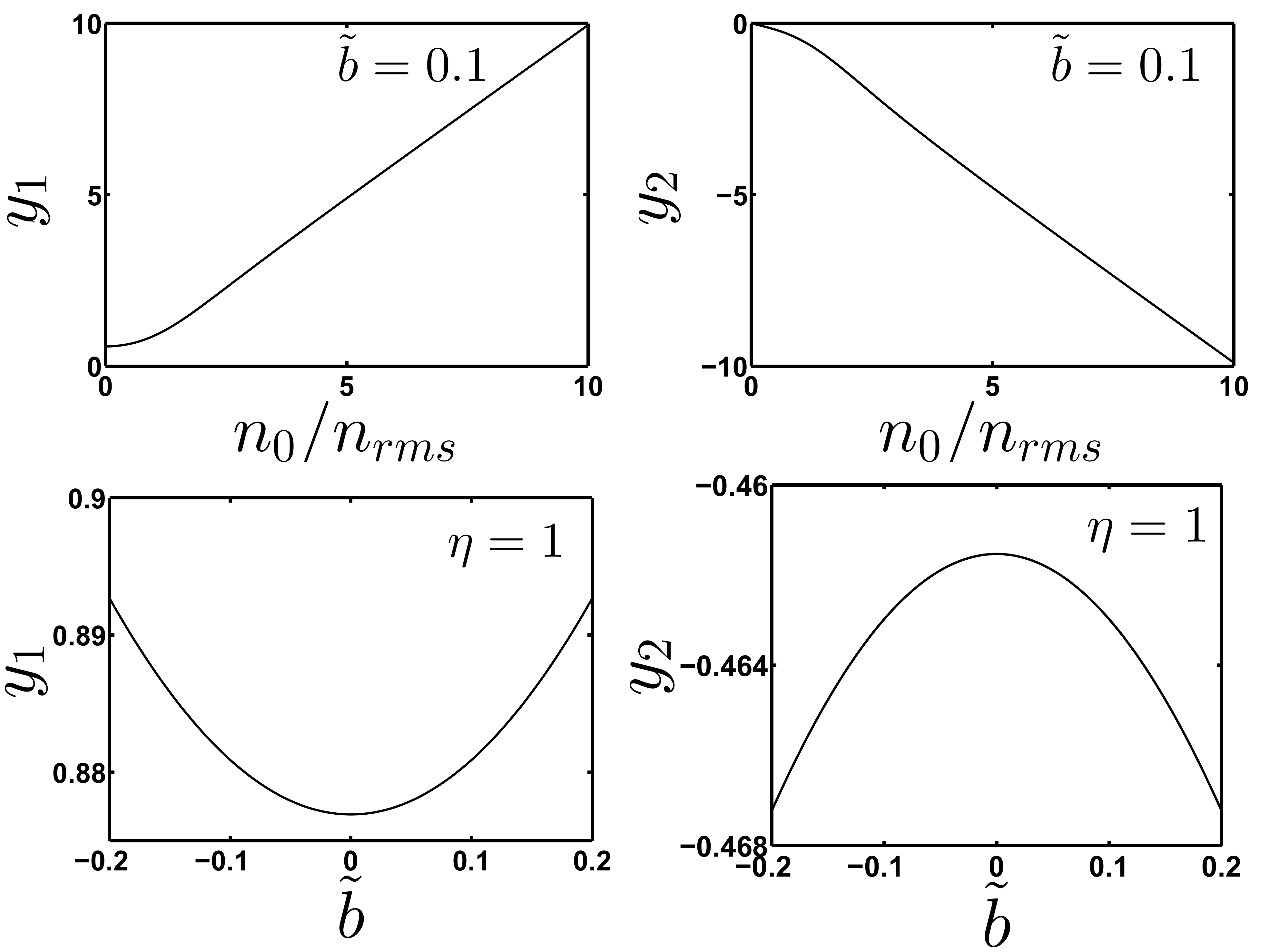}
\caption{\label{Fig:S2}Dependence of dimensionless
functions $y_1$ and $y_2$ on dimensionless magnetic field $\tilde{b}$ and $n_0/n_{\mathrm rms}$. The upper panel shows $y_1$ and $y_2$ as a function of $n_0/n_{\mathrm rms}$ at $\tilde{b}=0.1$ and the lower panel shows $y_1$ and $y_2$ as a function of $\tilde{b}$ at  $n_0/n_{\mathrm rms}=1$.}
\end{figure}

For the case where mobility $\mu$ is independent of carrier density, we introduce a dimensionless magnetic field ${\tilde b} = \mu B$ and dimensionless functions $y_1$ and $y_2$ such that $\sigma_{xx}^{{\rm EMT}}=y_{1}~n_{{\rm rms}}e\mu/\left[1+(\mu B)^{2}\right]$ and $\sigma_{xy}^{{\rm EMT}}=y_{2}~n_{{\rm rms}}e\mu^{2}B/\left[1+(\mu B)^{2}\right]$. Equation 3 of the main text then simplifies to
\begin{subequations}
\begin{equation}
\int_{-\infty}^{\infty} dy \ e^{-(y - \eta)^2/2} \frac{y^2 - y_1^2 + {\tilde b}^2 (y_2 - y)^2 }{(|y| + y_1)^2 + {\tilde b}^2 (y_2 - y)^2} = 0
\end{equation}
\begin{equation}
\int_{-\infty}^{\infty} dy \ e^{-(y - \eta)^2/2}  \frac{y_2 -y }{(|y| + y_1)^2 + {\tilde b}^2 (y_2 - y)^2} = 0.
\end{equation}
\end{subequations}
These equations can be easily solved numerically both for constant $\tilde{b}$ and also for constant $n_0/n_{\mathrm rms}$. Shown in Fig.~\ref{Fig:S2} are the dependence of dimensionless constants $y_1$ and $y_2$ on $\tilde{b}$ and $n_0/n_{\mathrm rms}$.
 
From the data in Fig.~\ref{Fig:S2}, we can obtain the coefficient of quadratic magnetoresistance through
\begin{equation}
A=1-\left[\left(\frac{y_{2}(0)}{y_{1}(0)}\right)^{2}+\frac{1}{2\mu^{2}}\frac{\partial_{B}^{2}y_{1}(0)}{y_{1}(0)}\right]
\end{equation}

\subsection{Non-constant mobility}

In reality, graphene has a mobility that is weakly dependent on carrier density due to the present of short-range scatterers $\sigma_{\mathrm{s}}$ and finite substrate distance $d$. In this case, we need to employ the full EMT equations of Eq. 3 of the main text. $\sigma_{B}[n_{0}]$ and $\mu_{B}[n_{0}]$ enter the EMT equation through longitudinal and transverse conductivity of the single channel model, as shown below.
\begin{equation}
\sigma[n_{0},B]=\frac{\sigma_{B}[n_{0}]}{1+(\mu_{B}[n_{0}]B)^{2}}\begin{pmatrix}1 & \pm\mu_{B}[n_{0}]B\\
\mp\mu_{B}[n_{0}]B & 1
\end{pmatrix}
\end{equation}
This is essentially Eq. 1 of the main text, except that we have the RPA--Boltzmann conductivity. $\sigma_{B}[n_{0}]=|n_{0}|e\mu_{B}[n_{0}]$ instead of $n_{0}e\mu$.

\begin{figure}
\includegraphics[width=3.2in]{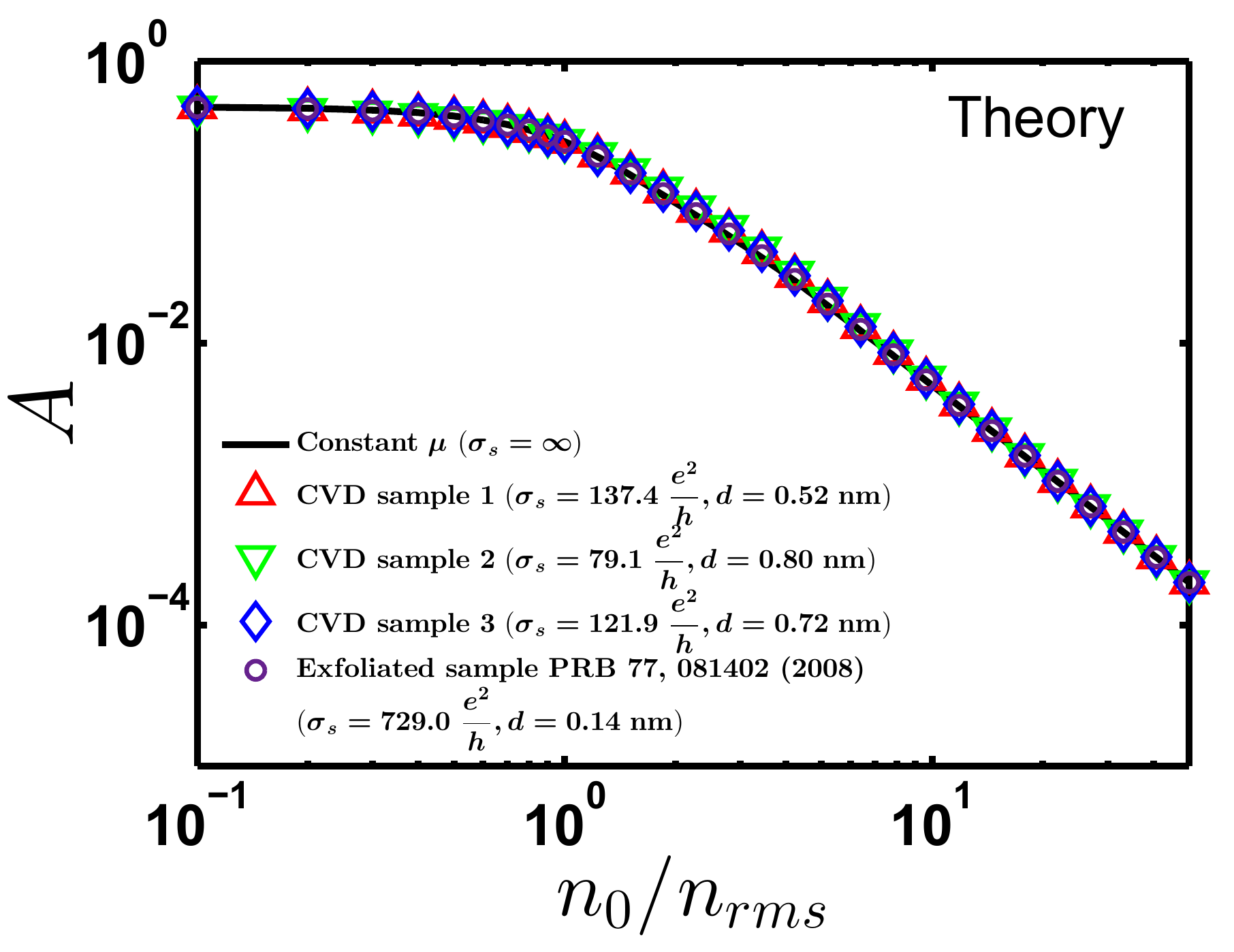}
\caption{\label{Fig:S3}(Color online) Theoretical results for the dependence of the coefficient of quadratic magnetoresistance $A$, on the ratio of carrier density $n_{0}$ and carrier density fluctuations $n_{\mathrm{rms}}$ for $\mu$ independent on carrier density and $\mu$ of the graphene samples, where it is weakly dependent on carrier density. $A$ determines the magnetoresistance through $\rho_{xx}[n_{0,}B]=\rho_{xx}(0)\left[1+A(\mu[n_{0}]B)^{2}\right]$. We find that the non-constant $\mu$ hardly affects $A$, and thus the power law dependence $A=(1/2)(n_{0}/n_{\mathrm{rms}})^{-2}$ still holds for $n\gg n_{\mathrm{rms}}$ for all samples.}
\end{figure}

We calculate the theoretical values of coefficient of the quadratic magnetoresistance $A$ of these samples. We find that the density dependent carrier mobility hardly affects $A$, as shown in Fig.~\ref{Fig:S3}. Since the density dependence of $\mu$ does not change the result, we use the simple density independent calculation of $A$ in Fig. 3 of the main text. Solving Eq. 3 of the main text numerically gives $\sigma_{xx}^{\mathrm{EMT}}$ and $\sigma_{xy}^{\mathrm{EMT}}$. The magnetoresistance and quadratic coefficient of magnetoresistance can then be obtained through
\begin{equation}
\rho_{xx}^{\mathrm{EMT}}=\frac{\sigma_{xx}^{\mathrm{EMT}}}{\left(\sigma_{xx}^{\mathrm{EMT}}\right)^{2}+\left(\sigma_{xy}^{\mathrm{EMT}}\right)^{2}}
\end{equation}
and
\begin{equation}
\label{Eq:rhoxxEMT}
\rho_{xx}^{\mathrm{EMT}}[n_{0,}B]=\rho_{xx}^{\mathrm{EMT}}(B=0)\left[1+A(\mu[n_{0}]B)^{2}\right]
\end{equation}

\subsection{Discrepancy of coefficient of quadratic magnetoresistance close to Dirac point}

Close to the Dirac point, we see that there are discrepancies between the theoretical and experimental values of the coefficient of quadratic magnetoresistance $A$. This discrepancy can be traced back to an overestimation of the value of self-consistent RPA--Boltzmann conductivity $\sigma_{B}$ (for small $\eta$) used in our model. To see this, we compare the theoretical and experimental values of $\sigma_{\mathrm{min}}$ at zero magnetic field and find that the theoretical $\sigma_{\mathrm{min}}$ in our model are larger than the experimental values by 20 \% to 45 \% for CVD-grown sample 1, sample 2, and sample 3. This overestimation of $\sigma_{B}$ is equivalent to an overestimation of mobility $\mu_{B}$ close to the Dirac point.

The lower experimental mobility close to the Dirac point is not captured by the fit value for $\mu[n_{0}]$ used in Eq.~\ref{Eq:rhoxxEMT}. This zero magnetic field effect can be attributed either to effect of island impurities found in the CVD-grown samples or to the finite resistance caused by p-n junctions between electron and hole puddles. We have checked that if we empirically modify $\mu[n_{0}]$ to account for scattering of p-n junctions and line defects, this discrepancy between theory and experiment at the Dirac point vanishes. However, this is just a phenomenological modification of $\mu[n_{0}]$ and a full theory of scattering by p-n junctions is beyond the scope of this work. In principle, we expect this discrepancy to disappear if one had a better way to measure $n_{\rm rms}$ directly in the experiment e.g. through scanning tunneling microscopy.

\subsection{Derivation of \texorpdfstring{$A\sim\left(\frac{n_{0}}{n_{{\rm rms}}}\right)^{-2}$}{}}
We can derive this result using both the schematic discussed in Fig. 1 of the main text and using a simplified effective medium theory. 

\subsubsection{Simple Physical Model}
Using the model established in Fig. 1 of the main text, we assume a magnetic field along the $z$ direction and an applied electric field along the $x$ direction. Examining two regions of electrons with carrier concentrations $n_{1}$ and $n_{2}$, we can relate the local currents to the local electric fields. For $i=1,2$ we
have 
\begin{equation}
\label{eq:conductivity}
\begin{pmatrix}J_{x}^{i}\\
J_{y}^{i}
\end{pmatrix}=\frac{n_{i}e\mu}{1+\mu^{2}B^{2}}\begin{pmatrix}1 & -\mu B\\
\mu B & 1
\end{pmatrix}\begin{pmatrix}E_{x}^{i}\\
E_{y}^{i}
\end{pmatrix}
\end{equation}

Noting that the steady state Hall field ensures no net current in the $y$ direction and assuming that the Hall field is the same in the two regions, we have
\begin{align}
E_{y} & =-\frac{\mu BJ_{x}}{n_{\rm avg}e\mu}\label{eq:Ey}
\end{align}

Enforcing current conservation in the $x$ direction, we have
\begin{align}
E_{x}^{i}=\frac{J_{x}}{n_{i}e\mu}+\frac{J_{x}\mu^{2}B^{2}}{e\mu}\left(\frac{1}{n_{i}}-\frac{1}{n_{\rm avg}}\right)\label{eq:Exfinal}
\end{align}

Assuming that the electric field across the sample may be redistributed over the regions to get different local field values but the net change in electric field due to the applied B field must be zero, we get
\begin{align}
\frac{J_{0}}{n_{1}e\mu}+\frac{J_{0}}{n_{2}e\mu} & =\frac{J_{x}}{n_{1}e\mu}+\frac{J_{x}\mu^{2}B^{2}}{e\mu}\left(\frac{1}{n_{1}}-\frac{1}{n_{\rm avg}}\right)\nonumber \\
 & +\frac{J_{x}}{n_{2}e\mu}+\frac{J_{x}\mu^{2}B^{2}}{e\mu}\left(\frac{1}{n_{2}}-\frac{1}{n_{\rm avg}}\right)
\end{align}

For $\mu B\ll 1$, 
\begin{equation}
J_{x}=J_{0}(1-\alpha^{2}\mu^{2}B^{2})
\end{equation}

We have defined $\alpha=\frac{(n_{1}-n_{2})}{(n_{1}+n_{2})}$. More
generally, we see that $A\sim\alpha^{2}\sim\left(\frac{\Delta n}{n_{\rm avg}}\right)^{2}\sim(\frac{n_{0}}{n_{{\rm rms}}})^{-2}$ as calculated in the full EMT model.

\subsubsection{Simplified effective medium theory}
The coefficient of quadratic magnetoresistance $A$ for $n_{0}\gg n_{\mathrm{rms}}$ can also be obtained from a one--band model for which the zero field conductivity is given by $\sigma_{0}=n_{0}e\mu$ with constant
$\mu$. Carrier density inhomogeneity is represented
by $P[n_{0}]=(1/2)\delta(n_{0}-n_{\mathrm{rms}})+(1/2)\delta(n_{0}+n_{\mathrm{rms}})$.
This model can be solved analytically for $|\eta|\gg1$ and $\mu B\ll1$. The longitudinal and transverse conductivities are given by 
\begin{align}
\sigma_{xx}^{\rm 2ch,EMT} & =\frac{|n_{0}|e\mu}{1+(\mu B)^{2}}\sqrt{\left(1-\eta^{-2}\right)\left[1+\eta^{-2}(\mu B)^{2}\right]}\label{eq:sigmaxx2chemt}\\
\sigma_{xy}^{\rm 2ch,EMT} & =-\frac{n_{0}\left(1-\eta^{-2}\right)e\mu}{1+(\mu B)^{2}}(\mu B)\label{eq:sigmaxy2chemt}
\end{align}
Thus, we have magnetoresistance
\begin{equation}
\rho_{xx}^{\rm 2ch,EMT} =\frac{\sqrt{1+\eta^{-2}(\mu B)^{2}}}{n_{0}\sqrt{1-\eta^{-2}}e\mu}
\end{equation}
which in turn gives
\begin{equation}\
A=\frac{1}{2}\eta^{-2}
\end{equation}
as what is obtained in the full effective medium theory.

\end{document}